\documentclass[a4paper,fleqn]{cas-dc}

\usepackage[numbers]{natbib}

\usepackage{url}
\usepackage{verbatim}
\usepackage{times}
\usepackage{epsfig}
\usepackage{graphicx}
\usepackage{amsmath,amsfonts,amssymb,amsthm,version}
\usepackage{algorithmic}
\usepackage{algorithm}
\usepackage{array}

\theoremstyle{}

\theoremstyle{remark}

\begin{document}
\let\WriteBookmarks\relax
\def\floatpagepagefraction{1}
\def\textpagefraction{.001}

\shorttitle{}    

\shortauthors{}  

\title [mode = title]{Robust Model Reconstruction Based on the Topological Understanding of Point Clouds Using Persistent Homology}  



\author[1]{Yu Chen}
\ead{chenyu.math@zju.edu.cn}

\author[1]{Hongwei Lin}
\cormark[1]
\ead{hwlin@zju.edu.cn}

\affiliation[1]{organization={School of Mathematics Science, Zhejiang University},
	city={Hangzhou},
	postcode={310058},
	country={China}}

\cortext[1]{Corresponding author}



\begin{abstract}
Reconstructing models from unorganized point clouds presents a significant challenge, especially when the models consist of multiple components represented by their surface point clouds. Such models often involve point clouds with noise that represent multiple closed surfaces with shared regions, making their automatic identification and separation inherently complex. In this paper, we propose an automatic method that uses the topological understanding provided by persistent homology, along with representative 2-cycles of persistent homology groups, to effectively distinguish and separate each closed surface. Furthermore, we employ Loop subdivision and least squares progressive iterative approximation (LSPIA) techniques to generate high-quality final surfaces and achieve complete model reconstruction. Our method is robust to noise in the point cloud, making it suitable for reconstructing models from such data. Experimental results demonstrate the effectiveness of our approach and highlight its potential for practical applications.
\end{abstract}

\begin{keywords}
surface reconstruction \sep persistent homology \sep representative cycle \sep  topological data analysis.
\end{keywords}

\let\printorcid\relax
\maketitle

\section{Introduction}
Surface reconstruction from unorganized point clouds represents a foundational challenge in computer graphics and geometric modeling, with broad applications spanning computer-aided design (CAD), medical imaging, computer animation, and virtual reality. The objective of this task is to reconstruct surfaces represented by point clouds that may contain noise, enabling downstream applications that require high-quality shapes. 

\begin{figure*}
	\centering
	\includegraphics[width=0.95\linewidth]{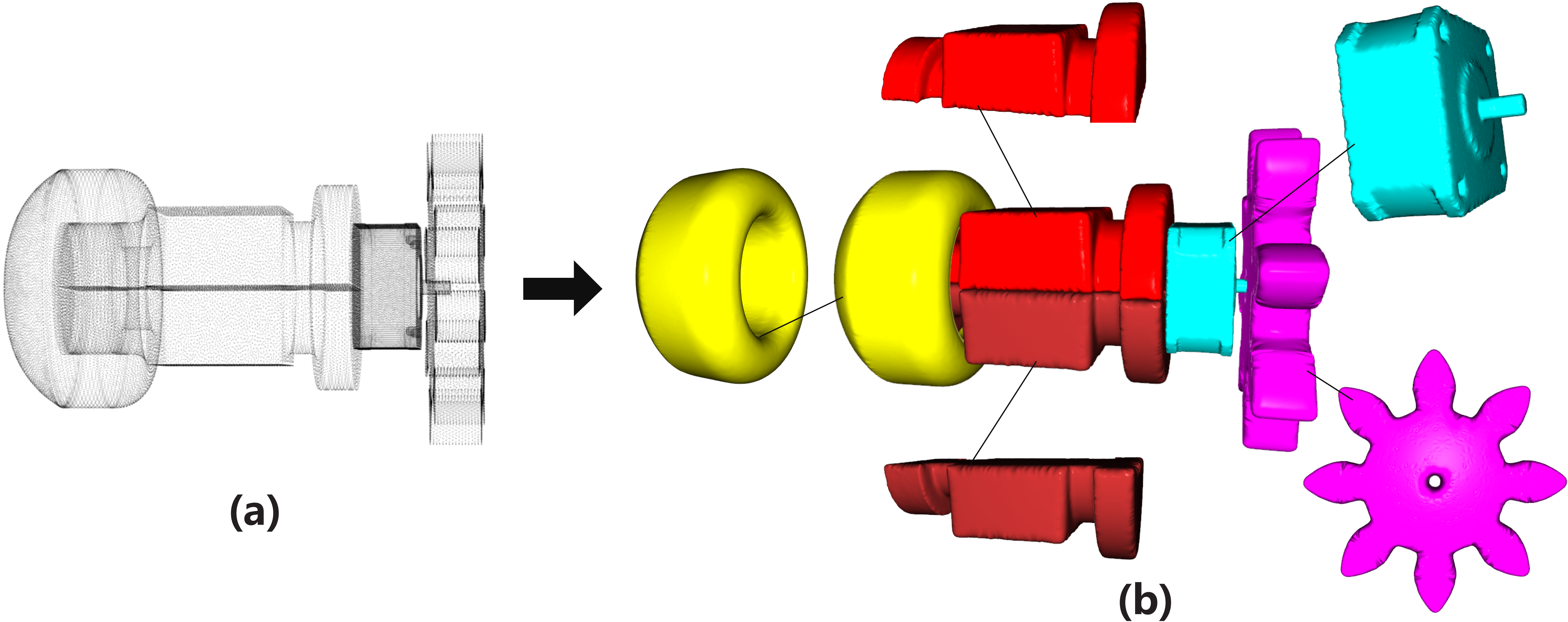}
	\caption{Reconstruct a whole mechanical model with several components from point cloud by our method. (a) A point cloud representing a composite mechanism composed of multiple components with shared regions. (b) All components are separated and reconstructed, each visualized in distinct colors.}
	\label{fig:robothand}
\end{figure*}

Traditional reconstruction methods usually focus on single surface reconstruction. When the object to be reconstructed is a complex model composed of multiple components, which may overlap or be nested within one another (see Fig. \ref{fig:robothand} for an example), traditional surface reconstruction methods face significant challenges when reconstructing the whole model directly from point clouds. This is particularly true when the given point cloud represents such a model, where the surface point clouds of individual components collectively form the input data. The presence of shared regions, nested structures, and noise further complicates the reconstruction process, making it difficult to achieve accurate and robust results. 

In such cases, distinguishing the points corresponding to each component's surface and reconstructing the entire model is a difficult task. These challenges, however, are of considerable practical importance. For example, Fig. \ref{fig:robothand} illustrates the decomposition of an industrial mechanism model, where the point clouds of all its components' surfaces are available. This point cloud represents multiple closed surfaces with shared regions, making the decomposition and accurate reconstruction of each component's surface both critically important and highly challenging.

\begin{figure*}
	\centering
	\includegraphics[width=1.0\linewidth]{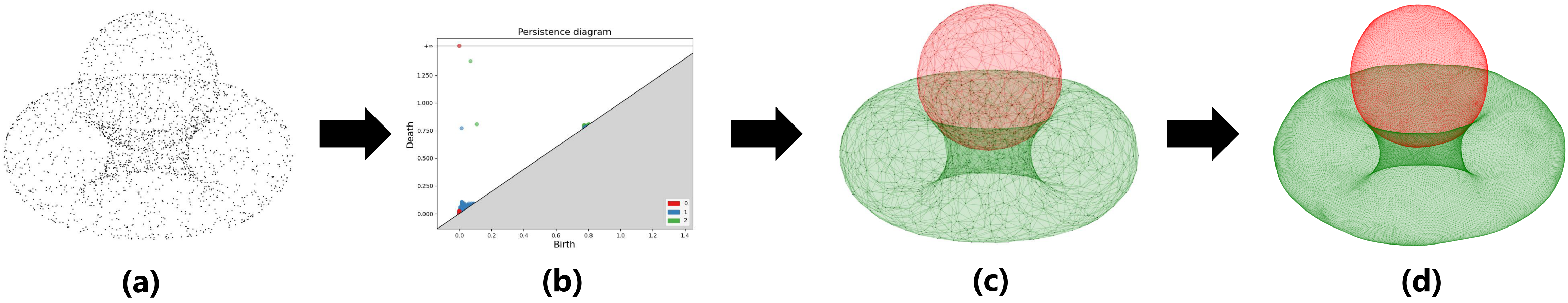}
	\caption{Flowchart of the proposed method. (a) The input point cloud (which may contain noise). (b) Topological understanding by computing 2-dimensional persistence diagrams. (c) Identify representative 2-cycles. (d) Loop subdivision fitting and LSPIA methods are applied to reconstruct the surfaces.}
	\label{fig:pipline}
\end{figure*}

In this paper, we develop a novel method for automatically reconstructing closed surfaces with sharing regions from unorganized point clouds by integrating topological analysis with geometric processing techniques, and address the challenges of reconstructing multi-component models from point clouds with noise. Specifically, we first employ persistent homology to gain a topological understanding of the point cloud, which is robust to noise and can effectively identify the number of surfaces present for reconstruction. Subsequently, based on the additional information provided by persistent homology, representative cycles are used to approximate the surfaces, yielding initial triangular mesh representations. Finally, the Loop subdivision \cite{loop1987smooth} fitting with LSPIA \cite{deng2014progressive}, is applied to generate high-quality reconstructed surfaces, effectively addressing the challenge of reconstructing surfaces of models with multiple components and common regions.
Fig. \ref{fig:pipline} shows a flowchart of the entire surface reconstruction procedure. 

In summary, the main contributions of this study are as follows:
\begin{itemize}
	\item The number of reconstructed closed surfaces is determined by topological understanding using persistent homology.
	\item The initial control meshes for surface reconstruction are represented by the representative cycles of persistent homology.
	\item The LSPIA method for subdivision surfaces is employed to generate reconstructed surfaces that better approximate the given point clouds.
\end{itemize}

The remainder of this paper is organized as follows. Section \ref{sec:RW} reviews related work on surface reconstruction and computational topology. Section \ref{sec:Preliminaries} introduces the preliminaries on persistent homology and representative cycles. In Section \ref{sec:Algorithm}, we present our proposed approach for reconstructing models from point clouds containing multiple closed surfaces. Experimental results demonstrating the effectiveness of our method and some discussions are provided in Section \ref{sec:exp}. Finally, Section \ref{sec:conclusion} concludes the study.

\section{Related Work}\label{sec:RW}
\subsection{Surface reconstruction methods}
Surface reconstruction from point clouds remains a fundamental challenge in computer graphics, with applications spanning virtual reality, robotics, and industrial design \cite{huang2024surface}. Existing methods can be categorized into four main paradigms, each with distinct advantages and limitations. In the following, we survey some literature closely related to our work. For more details on surface reconstruction, please refer to the survey paper \cite{huang2024surface}.

Triangulation-based methods construct surfaces by connecting points through geometric primitives like triangles, especially the Delaunay-based reconstruction methods \cite{cazals2006delaunay}. The Greedy Delaunay algorithm \cite{cohen2004greedy} iteratively selects optimal triangles from Delaunay candidates, while the Ball-Pivoting Algorithm (BPA) \cite{bernardini1999ball} generates meshes by rolling spheres over point neighborhoods. Weighted alpha shapes \cite{edelsbrunner1992weighted} is also a powerful and common tool for triangulation-based reconstruction using alpha complexes. Additionally, \cite{cazals2012reconstructing} proposed a method to reconstruct more general shapes (compact sets), which can handle noise and boundaries. 
The IPD algorithm \cite{lin2004mesh} constructs a triangular mesh by incrementally adding new points starting from seed triangles to achieve an approximate minimum weight triangulation reconstruction of the surfaces.
When the point cloud satisfies the $\varepsilon$-sample condition \cite{amenta1998surface}, these methods come with precise theoretical guarantees on the topology and geometry of the reconstructed surface. However, some of their performance is particularly sensitive to parameter selection, and may yields fragmented reconstructions when point density varies or outliers are present.

Smoothness-prior methods regularize surfaces through global optimization. Poisson Surface Reconstruction (PSR) \cite{kazhdan2006poisson} solves an implicit function constrained by oriented normals, while its screened variant \cite{kazhdan2013screened} adds spatial adaptivity. Algebraic Point Set Surfaces \cite{guennebaud2007algebraic} and Robust Implicit Moving Least Squares (RIMLS) \cite{oztireli2009feature} extend moving least squares with curvature-aware weighting. Though robust to moderate noise, these methods over smooth fine details and require accurate normal estimation. They fail catastrophically on misaligned multi-view scans and cannot handle large missing regions.

Template-based methods leverage geometric priors through primitive fitting or model retrieval. RANSAC-based approaches \cite{schnabel2007efficient} iteratively fit geometric primitives, while retrieval methods \cite{pauly2005example} deform pre-existing CAD models. Hybrid techniques like Delaunay Surface Elements (DSE) \cite{rakotosaona2021learning} combine Delaunay triangulation with learned feature descriptors. While effective for structured environments, these methods lack generality. RANSAC struggles with specific shapes, and retrieval approaches require extensive template libraries. They also fail to reconstruct surfaces deviating significantly from template geometries.

Deep learning methods learn implicit or explicit shape representations. Global implicit models like DeepSDF \cite{park2019deepsdf} encode shapes in latent spaces, whereas local approaches like LIG \cite{jiang2020local} partition space into voxels with independent codes. Points2Surf \cite{erler2020points2surf} learns features from both local patches and the global surface, and reconstructs the surface with an implicit decoder. Transformer-based architectures \cite{yan2022shapeformer} enhance detail preservation through attention mechanisms. Despite noise robustness, these methods require massive training data and generalize poorly to unseen categories. As shown in \cite{huang2024surface}, they underperform classical methods on complex real-world scans with misalignment or severe missing data, due to their reliance on learned shape priors.

Furthermore, these existing methods have difficulty in reconstructing the surfaces of individual components of a combined model from a point cloud directly. In this work, we aim to leverage the topology representative cycle and the subdivision method to accurately identify and effectively reconstruct closed surfaces within point clouds, ensuring robust reconstruction of the surfaces of all components of a combined model.

\subsection{Computational topology}
Computational topology, particularly persistent homology, has emerged as a powerful framework for analyzing the topological features of data \cite{edelsbrunner2002topological,Carlsson2009TopologyAD,edelsbrunner2008persistent,edelsbrunner2022computational}. Persistent homology enables the identification of topological features on multiple scales, making it particularly effective for analyzing noisy data. As a foundation of topological data analysis (TDA), it extracts topological features from point clouds through a filtration process, facilitating the detection of structures such as connected components, loops and holes \cite{carlsson2014topological}. Previous research has used persistent homology for tasks such as surface feature extraction \cite{dey2008computing}, surface optimization \cite{bruel2020topology,dong2022topology}, and curve reconstruction \cite{he2023robust,chen2025human}. However, few relative methods have been studied to tackle the challenge of reconstructing multiple surfaces with holes from unorganized point clouds using topological insights. 

In this work, we utilize persistent homology to provide a topological understanding of point clouds, offering critical insights for surface reconstruction. Additionally, we use the representative cycles of persistent homology to get the initial approximation of surfaces to be reconstructed.

\section{Preliminaries}\label{sec:Preliminaries}
In this section, we introduce the key concepts that are central to our method.

\subsection{Alpha filtration}
An \textit{$n$-simplex} is defined as the convex hull formed by $n+1$ affinely independent points $\left\lbrace u_0, u_1, \ldots, u_n \right\rbrace$ in the Euclidean space $\mathbb{R}^N$, denoted as $ \left[u_0, u_1, \ldots, u_n \right] $. Geometrically, an $n$-simplex can be represented as a geometric model, such as vertex (0-simplex), line segment (1-simplex), triangle (2-simplex), and tetrahedron (3-simplex). 
A \textit{simplicial complex} $K$ is a collection of simplices that satisfies specific properties: every face of a simplex in $K$ is also in $K$, and the intersection of any two simplices in $K$ is a face of both of them \cite{edelsbrunner2022computational}. The dimension of a simplicial complex $K$ is defined as the maximum dimension among all the simplices in $K$. A \textit{subcomplex} of $K$ is a simplicial complex $L \subseteq K$.

An \textit{alpha complex} is a special type of simplicial complex. To illustrate it, we first introduce the \textit{Delaunay complex} derived from a point cloud. Let $P=\{ x_i\}_{i=1}^m$ be a point cloud in $\mathbb{R}^n$. For each point $x_i$, we define its \textit{Voronoi cell} as 
$$V_{x_i} = \{ x\in \mathbb{R}^n: \|x-x_i\|\le \|x-x_j\|, \forall j\ne i \},$$
where $\| \cdot\|$ denotes the Euclidean norm in $\mathbb{R}^n$. The \textit{Delaunay complex} generated by $P$ is given by 
$$\operatorname{Del}(P)=\left\{ \left[x_{i_{1}}, \cdots, x_{i_{q}}\right] : V(x_{i_{1}}) \cap \cdots \cap V(x_{i_{q}}) \neq \emptyset\right\}.$$

An alpha complex generated by a set of points $P$ is a subcomplex of the Delaunay complex, parameterized by a non-negative value $r$. For a point $ x \in P$, let $ B_x(r) $ denote the closed ball centered at $ x $ with radius $ r $. Furthermore, define $R_x(r) = B_x(r) \cap V_x$, where $V_x$ represents the Voronoi cell associated with $x$. The alpha complex generated by $P$ is then defined as
$$
\operatorname{Alp}(P, r) = \left\{ \left[x_{i_{1}}, \dots, x_{i_{q}}\right] : R_{x_{i_{1}}}(r) \cap \cdots \cap R_{x_{i_{q}}}(r) \neq \emptyset \right\}.
$$
Since $R_x(r) \subseteq V_x$, it follows that the alpha complex is a subcomplex of the corresponding Delaunay complex.

\begin{figure*}
	\centering
	\includegraphics[width=1.0\linewidth]{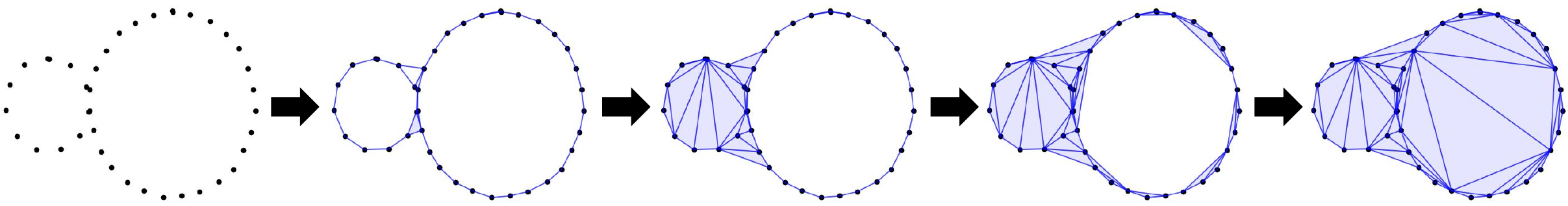}
	\caption{An example of alpha filtration. As the filtration parameter increases, more and more simplices appear.}
	\label{fig:alpha}
\end{figure*}

Note that each alpha complex is associated with a free parameter $r$, for a fixed point cloud $P\in \mathbb{R}^n$, if $r$ is gradually increased starting from 0, a sequence of complexes is obtained, satisfying
$$\emptyset = \operatorname{Alp}(P, r_0) \subset \operatorname{Alp}(P, r_1) \subset \cdots \subset \operatorname{Alp}(P, r_k)\subset \cdots.$$
This sequence is referred to as the \textit{alpha filtration} of $P$. An example of alpha filtration is illustrated in Fig. \ref{fig:alpha}.

\subsection{Persistent homology and representative cycle}
Persistent homology is a tool from computational topology that captures the topological features of space at various scales \cite{edelsbrunner2008persistent,edelsbrunner2022computational}. It provides a way to quantify the persistence of topological features, such as connected components, loops and holes. 

Let $K$ be a simplicial complex and $n$ its dimension. An \textit{$n$-chain} is a sum of $n$-simplices in $K$, denoted by $c=\sum a_i\sigma_i$, where $\sigma_i$ represents an $n$-simplex, and here $a_i$ is its coefficient belonging to $\mathbb{Z}_2$. With the addition operation, the $n$-chains form the $n$-chain group, denoted by $C_n(K)$.
The \textit{boundary} for a $n$-simplex $\sigma=[x_0,x_1,\ldots,x_n]$ is given by
$$
\partial_{n}\sigma=\sum_{j=0}^{n} [x_{0},\ldots,\hat{x}_{j},\ldots,x_{n}],
$$
where $\hat{x}_j$ denotes the omission of $x_j$. Let $Z_n(K)=\operatorname{Ker}(\partial_n)$ be the group of \textit{$n$-cycle} and $B_n(K)=\operatorname{Im}(\partial_{n+1})$ the group of \textit{$n$-boundary}. The \textit{$n$-th homology group} of $K$ is defined as the quotient group $H_{n}(K)=Z_{n}(K)/B_{n}(K)$.

Now suppose we have a filtration (in this study alpha filtration will be considered)
$$K_0\subset K_1\subset \cdots \subset K_m,$$
then for every $0\leq i\leq j\leq m$ we have an inclusion map from $K_i$ to $K_j$ and therefore an induced homomorphism $f_n^{i,j}: H_n(K_i)\to H_n(K_j)$ for each dimension $n$. The $n$-th \textbf{persistent homology groups} are defined as the images of the homomorphisms induced by inclusion:
$$H_n^{i,j}=\operatorname{Im}f_n^{i,j},\: \forall  0\leq i\leq j\leq m.$$ 
The corresponding $n$-th \textbf{persistent Betti numbers} are the ranks of these groups: $\beta_{n}^{i,j}=\operatorname{rank}H_{n}^{i,j}$.
From the definition, it is evident that persistent homology captures the variation (birth and death times) of cycles within the filtration, thereby revealing the topological features of the point cloud.

\begin{figure}
	\centering
	\includegraphics[width=1.0\linewidth]{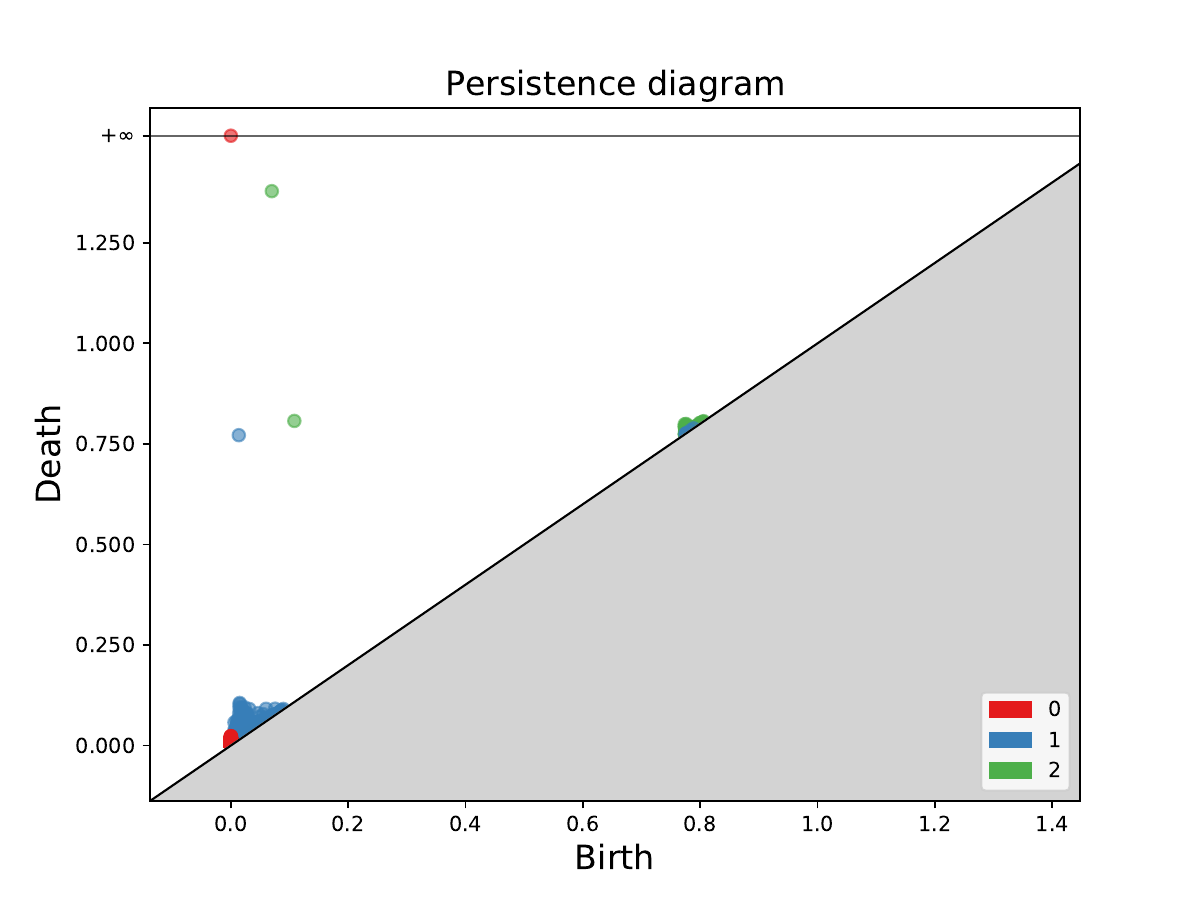}
	\caption{An example of a persistence diagram. The red, blue, and green points correspond to the 0-th, 1-st, and 2nd persistent homology groups respectively.}
	\label{fig:PD}
\end{figure}

A useful tool for visualizing persistent homology is the \textit{persistence diagram} (PD) \cite{edelsbrunner2022computational}, as shown in Fig. \ref{fig:PD}.
The set of points that record the birth time and death time of $n$-cycles is called the $n$-PD. 
Each point in a $n$ -PD is represented as $(b_i,d_i)$, corresponding to an $n$-cycle and capturing its \textit{birth time} $b_i$ and \textit{death time} $d_i$. 
The value $\left| b_i-d_i\right| $ is defined as the \textit{persistence} of this cycle.
Additionally, consider a filtration that introduces one simplex at a time (an alpha filtration satisfies this condition in practice). We define a simplex $\sigma$ as \textit{positive} if its addition creates a cycle, thereby giving birth to a new homology class. Otherwise, we define it as \textit{negative}. 
For a 2-cycle in an alpha filtration, the positive simplex is a 2-simplex, while the negative simplex is a 3-simplex.
Actually, for a point in $n$-PD and its corresponding $n$-cycle, the birth time indicates when the positive simplex appears in the filtration, while the death time indicates when the negative simplex appears in the filtration.

Generally, points within an $n$-PD can be categorized based on their persistence. Points with higher persistence are termed \textit{significant points}, whereas those with lower persistence are referred to as \textit{noise points}. Similarly, a $n$-cycle corresponds to a significant point is a \textit{significant $n$-cycle}; otherwise, it is a \textit{noise $n$-cycle}. 
In this study, we focus on closed surfaces within point clouds, which correspond to 2-cycles. Therefore, we will identify significant points in 2-PD and track the birth and death times of significant 2-cycles in the alpha filtration to make a topological understanding of the point cloud.

For each homology group $H_n(K)$, suppose its basis is $\left\lbrace [r_1],\cdots,[r_k] \right\rbrace $, then each $r_i \: (i=1,\cdots,k)$ is referred to as a \textit{representative $n$-cycle} of the homology class $[r_i]$. In the context of persistent homology, the representative $n$-cycle of a persistent pair $(b,d)$, known as a \textit{persistent $n$-cycle} \cite{dey2019persistent}, is a specific representative $n$-cycle in the filtration $K^b\subset K^{b+1}\subset \cdots \subset K^d$. For instance, a persistent $n$-cycle associated with $(b,d)$ first emerges at $K^b$ and becomes a boundary at $K^d$. 
In our method, we adopt the \textit{volume optimal cycle} \cite{obayashi2018volume} as the representative cycle for each point in the 2-PD.

In this study, since a representative 2-cycle corresponds to a closed triangular surface, we utilize Loop subdivision \cite{loop1987smooth} to enhance the quality of the reconstructed surface. A key challenge in applying Loop subdivision to a point cloud lies in obtaining a high-quality initial control mesh from the point cloud, and representative 2-cycles effectively address this issue. Furthermore, this method preserves the overall shape of the surface while iteratively subdividing triangles into smaller ones, thereby improving the resolution of the mesh.

\section{Reconstructing Models Represented by Multiple Closed Surfaces}\label{sec:Algorithm}
In this section, we propose an algorithm for reconstructing closed surfaces from a point cloud using persistent homology, Loop subdivision, and optimization techniques (LSPIA). The point cloud, representing a model, may encompass multiple closed surfaces to be reconstructed, which could share common regions, and may also contain noise. 
We assume the point clouds are sampled sufficiently, which means all regions on surfaces should be sampled relatively uniformly, and no region can be left unsampled. For example, assuming the point cloud satisfies the $\varepsilon$-sample condition with sufficiently small $\varepsilon$ \cite{amenta1998surface}, which can be satisfied in most of the practical situations.

\subsection{Topological understanding of the point cloud}
If the given point cloud represents multiple closed surfaces with shared common regions, such as two tangent spheres and a torus, or two adjacent tetrahedrons on a common face, it becomes essential to identify and reconstruct each closed surface individually. Persistent homology provides a powerful tool for distinguishing and separating all such closed surfaces that need to be reconstructed.
Based on the 2-PD derived from the alpha filtration of the original point cloud, we can infer the 2-dimensional topological features of the point cloud. Specifically, the number of closed surfaces that can be reconstructed from the point cloud can be determined by analyzing the corresponding 2-PD. However, due to the presence of noise in the original point cloud, numerous noise points may appear in the 2-PD, corresponding to small-scale 2-cycles. To eliminate these spurious cycles and extract significant topological features, we will focus on the significant points in the 2-PD.

\begin{figure}
	\centering
	\includegraphics[width=0.9\linewidth]{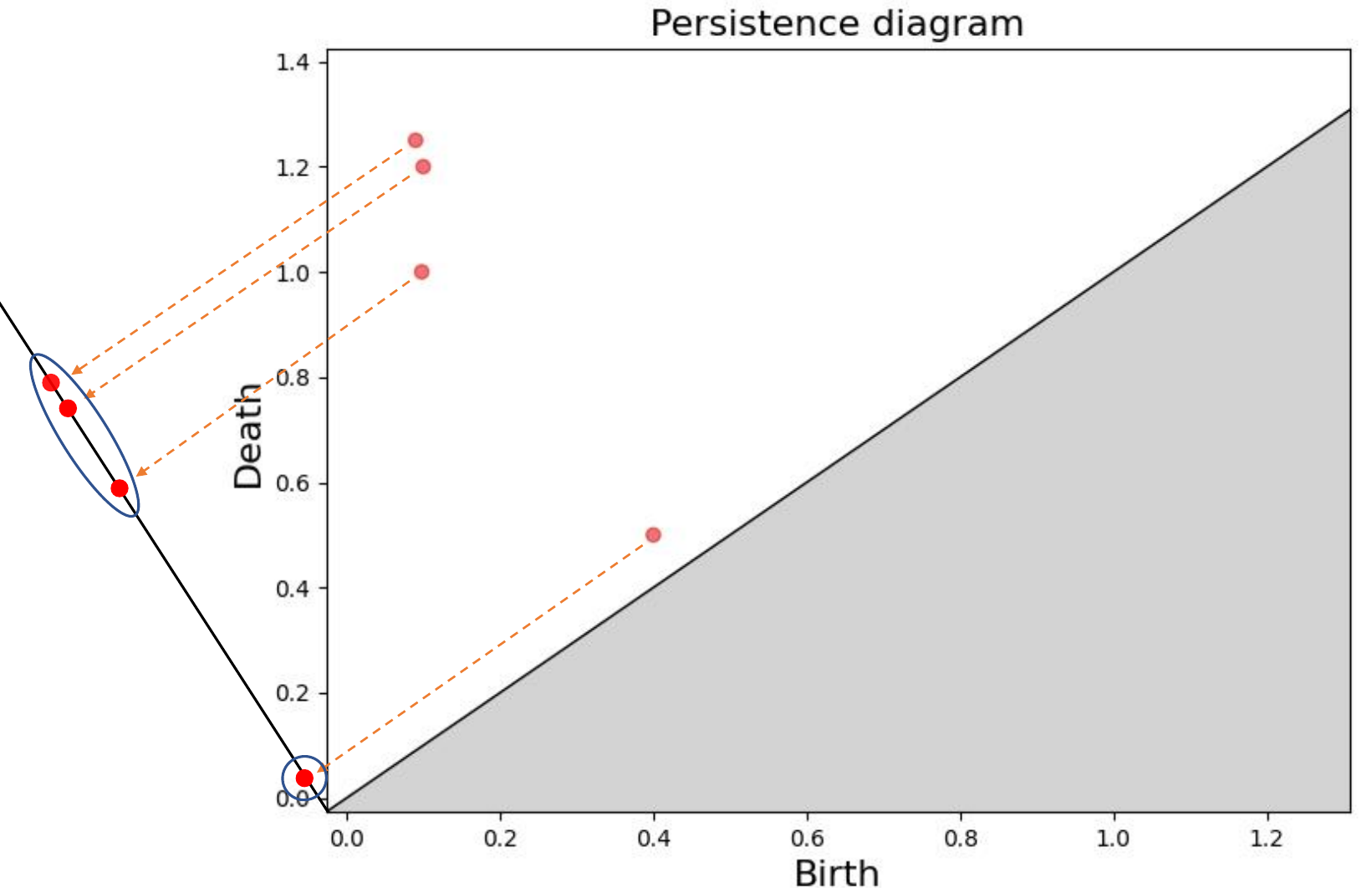}
	\caption{An example of clustering points in a PD.}
	\label{fig:PDclustering}
\end{figure}

To identify significant points in 2-PD, we initially project the points in 2-PD onto the line $y=-x$. The distance from these projected points to the origin $(0,0)$ serves as a proxy for the persistence of the corresponding points in the 2-PD. The projected points are then divided into two classes using the $k$-means clustering algorithm (in our method $k=2$, since in the analysis of the persistence diagram, the common practice is to divide the points inside into two classes, i.e. significant points and noise points) \cite{hamerly2002alternatives}. Points belonging to the class with greater persistence are regarded as significant, and each significant point corresponds to a closed surface that is to be reconstructed from the point cloud. Fig. \ref{fig:PDclustering} illustrates this clustering process of a PD.

\subsection{Computing representative 2-cycles}
From the topological understanding of point cloud, we get the number of significant points in 2-PD, hence the number of closed surfaces to be reconstructed.
Subsequently, the \textit{volume optimal cycle} \cite{obayashi2018volume} is adopted in our implementation, which will be used as the representative cycle for each significant point in the 2-PD. For a point $(b_i,d_i)$ in the 2-PD, it corresponds to a 2-cycle that initially forms at time $b_i$ and finally becomes the boundary of a 3-chain at time $d_i$. Consequently, we can identify this 3-chain first, and then compute its boundary to obtain the 2-cycle that corresponds to $(b_i,d_i)$ in the 2-PD.
Define 
$$\mathcal{F}_{q}=\{\sigma :q\text{-simplex}, \sigma \in \operatorname{Alp}(P,d_i)-\operatorname{Alp}(P,b_i)\},$$
and notice that this 3-chain is a set of 3-simplex from $\mathcal{F}_{3}$, and contains the negative 3-simplex $\sigma_{d_{i}}$ of point $(b_i,d_i)$. Therefore, we can find this 3-chain $z$ by solving the following optimization problem \cite{obayashi2018volume}:
\begin{align}
	& \text{minimize } \|z\|_{1} \text{ subject to } \nonumber \\
	& z = \sigma_{d_{i}} + \sum_{\sigma_{k} \in \mathcal{F}_{3},\: \alpha_k \in \mathbb{Z}_2} \alpha_{k} \sigma_{k}, \label{eq:1} \\
	& \tau^{*}(\partial z) = 0, \quad \forall \tau \in \mathcal{F}_{2} \label{eq:2} \\
	& \sigma_{b_i}^{*}(\partial z) \ne 0. \label{eq:3}
\end{align}
where $\|z\|_{1}=1+\sum |\alpha_k|$, and $\sigma_{i}^{*}$  is the linear map on $C_{2}$ satisfying $ \sigma_{i}^{*}(\sigma_{j})=\delta_{i j}$ (Kronecker symbol) for any 2-simplex $ \sigma_{i}, \sigma_{j}$. The optimal solution $\hat{z}$ is called the \textit{persistent volume} for $(b_i,d_i)$, and its boundary $\partial \hat{z}$ is called the \textit{volume-optimal cycle} for $(b_i,d_i)$. 
In fact, the restricted conditions imply that the persistent volume associated with the pair $(b_i,d_i)$ must include the negative 3-simplex $\sigma_{d_i}$, while other 3-simplices should be selected from those that first appear within the period between $b_i$ and $d_i$. Moreover, the corresponding volume-optimal cycle must contain the positive 2-simplex $\sigma_{b_i}$ but exclude 2-simplices that appear during the period between $b_i$ and $d_i$. Consequently, the volume-optimal cycle is consistent with the information provided by persistent homology: a cycle emerges at time $b_i$, and it is finally filled by higher-dimensional simplices at time $d_i$.
Additionally, it has been proved that the optimization problem of the volume-optimal cycle always has a solution \cite{obayashi2018volume}. 

\begin{figure*}
	\centering
	\includegraphics[width=0.85\linewidth]{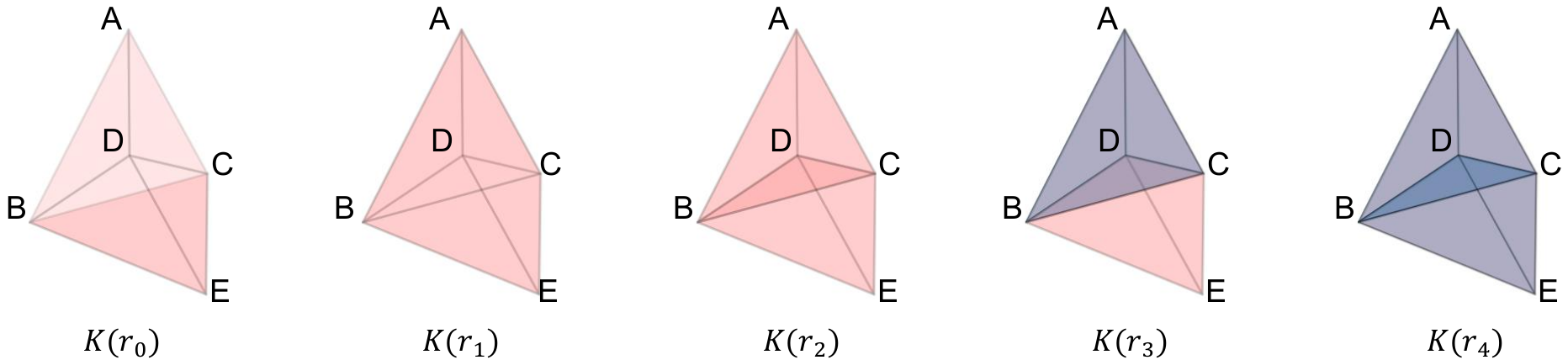}
	\caption{An example for illustrating the computation of volume-optimal cycle.}
	\label{fig:volopt}
\end{figure*}

Here we use Fig. \ref{fig:volopt} as an example of computing volume-optimal cycle. In this filtration, the simplicial complex $K(r_0)$ contains five 2-simplices: $[ABD]$, $[ACD]$, $[BCE]$, $[BDE]$, $[CDE]$ (shown in red). In $K(r_1)$, the 2-simplex $[ABC]$ is added, generating a 2-cycle: 
$$[ABC]+[ABD]+[ACD]+[BCE]+[BDE]+[CDE].$$
Then in $K(r_2)$, $[BCD]$ is added, while the 3-simplices $[ABCD]$ and $[BCDE]$ (shown in blue) are added in $K(r_3)$ and $K(r_4)$, respectively. We observe that in this filtration, there is a persistent pair $(r_1,r_4)$, where $[ABC]$ is the positive simplex and $[BCDE]$ is the negative simplex. We now compute the persistent volume and volume-optimal cycle for it. According to the definition, the persistent volume must contain $[BCDE]$, and thus it can be either $[BCDE]$ or $[ABCD]+[BCDE]$. However, $$\partial [BCDE]=[CDE]+[BDE]+[BCE]+[BCD],$$ 
and for $[BCD]\in K(r_4)-K(r_1)$,
$$[BCD]^* (\partial[BCDE])=[BCD]^* ([BCD])=1\ne 0.$$
Thus $[BCDE]$ does not satisfy the condition (Eq. \ref{eq:2}) and is not a persistent volume for $(r_1,r_4)$. On the other hand, we can similarly check that $[ABCD]+[BCDE]$ is a persistent volume for $(r_1,r_4)$, and its boundary
$$[ABC]+[ABD]+[ACD]+[BCE]+[BDE]+[CDE]$$
is the corresponding volume-optimal cycle.

Based on the preceding discussion, it is evident that the volume-optimal cycle serves as a geometrically meaningful and high-quality representative cycle for the significant points in the PD. It effectively captures the shape of the closed surfaces to be reconstructed from the point cloud and is well-suited for handling point clouds that contain multiple closed surfaces with common regions.

\subsection{Removing non-manifold vertices and edges}
\begin{figure}
	\centering
	\includegraphics[width=0.7\linewidth]{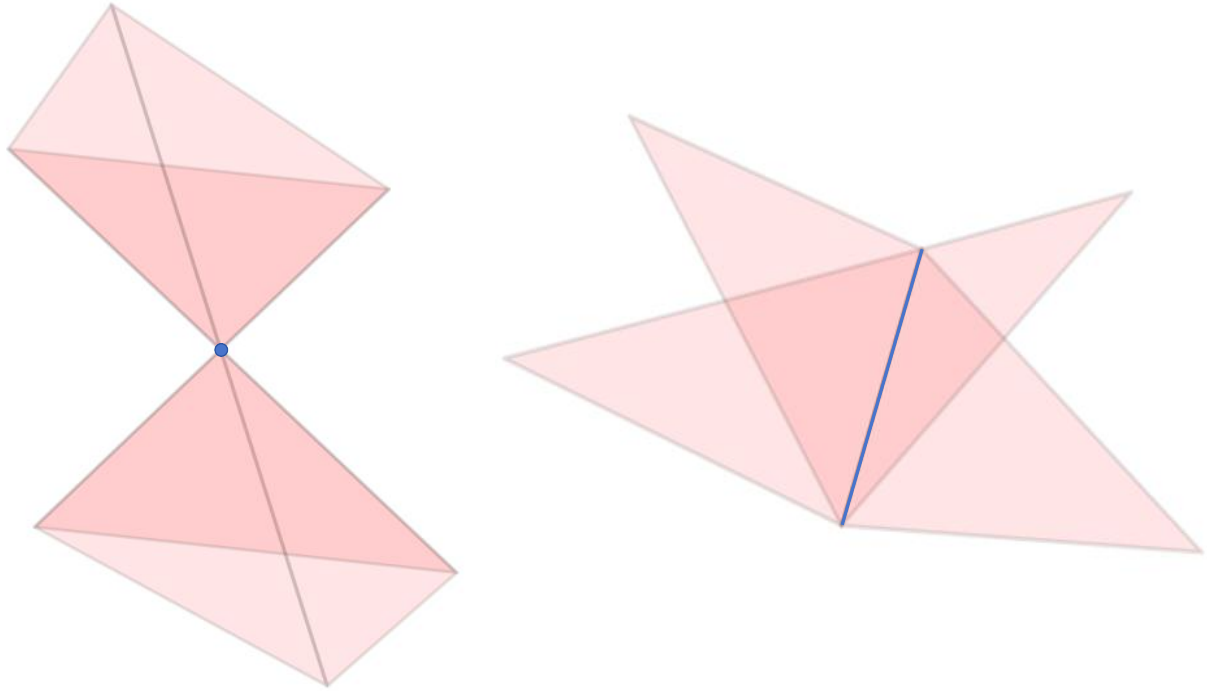}
	\caption{Left: an example of a non-manifold vertex.
		Right: an example of a non-manifold edge.}
	\label{fig:non-manifold}
\end{figure}

When a volume-optimal cycle is obtained, since the original point cloud may have noise, it may contain non-manifold vertices and edges (see Fig. \ref{fig:non-manifold} for examples). Since these vertices and edges will lead to wrong results of Loop subdivision, it is necessary to detect non-manifold vertices and edges and remove them. 

Suppose we have obtained a persistent volume $vol$ and the volume-optimal cycle $c$ corresponding to a significant 2-PD point. For each vertex $v$ in $c$, we first check the neighboring faces $F(v)$ of $v$. Starting from one face $F_1 \in F(v)$, we identify another face $F_2 \in F(v)$ that shares a common edge with $F_1$. This search process is repeated until $F_1$ is reencountered. If all faces in $F(v)$ are visited during this process, $v$ is classified as a normal vertex; otherwise, it is considered a non-manifold vertex. After finding all non-manifold vertices, we remove all 3-simplices in $vol$ adjacent to at least one non-manifold vertex, thus deriving an updated $vol$ and $c$. 

Next, we eliminate non-manifold edges. Since the volume-optimal cycle $c$ is the boundary of a combination of 3-simplices, we define non-manifold edges in $c$ as those connected to more than two faces. After identifying these edges and removing all their neighboring 3-simplices in $vol$, the cleaned persistent volume $vol^{new}$ will not contain non-manifold vertices or edges.
Hence, we finally compute $\hat{c} = \partial (vol^{new})$ again. The resulting $\hat{c}$ is now a representative 2-cycle without non-manifold vertices and edges, which is feasible for further optimization steps.

\subsection{Loop subdivision fitting by LSPIA}\label{LSPIA}
Although each 2-cycle forms a closed triangular mesh surface after removing non-manifold vertices and edges, it only comprises a subset of the points from the point cloud representing the corresponding sampled closed surface. Particularly in the presence of noise, the number of vertices in the representative 2-cycle may be significantly fewer than those in the point cloud representing the sampled closed surface, and the shape of the 2-cycle can be affected seriously by noise. As a result, these 2-cycles provide only a coarse approximation of the surface and cannot be directly used as the final reconstructed surfaces. To achieve higher-quality surfaces, the subsequent step involves applying Loop subdivision fitting to the obtained representative 2-cycles.

Initially, since there may be multiple 2-cycles in the point cloud, for each 2-cycle $c$ and their vertices $V_{c} = \{v_1, \dots, v_n\}$, it is necessary to identify a subset of the input point cloud in proximity to the 2-cycle for subsequent computations. We first compute the minimum distance $d_{min, i}$ from each data point $v_i$ to the remaining points in the input point cloud, followed by calculating the \textit{average distance} $d_{avg} = \frac{1}{n}\sum_{i=1}^n d_{min, i}$. This average distance serves as a threshold to identify points in $P$ that are close to each 2-cycle, effectively mitigating the influence of noise associated with the 2-cycle. Specifically, by defining the subset $P'_c$ of the original point cloud $P$ as 
\begin{align}
	P'_c = \{p \in P : \operatorname{dist}(p, V_{c}) \leq d_{avg}\}, \label{eq:4}
\end{align}
we obtain a collection of points sufficiently close to $c$, which can be utilized for further subdivision fitting.

Then we use $P'_c$ as target points for Loop subdivision fitting as follows.
Since the original point cloud may contain noise, to handle these challenges and get better approximate surfaces, we utilize Loop subdivision and \textit{least squares progressive iterative approximation} (LSPIA) method \cite{deng2014progressive} to optimize the obtained surfaces.
To handle cases of noise, we first reduce the number of meshes of each representative 2-cycle $c$ to around less than 1/4 using the topology-preserved QEM mesh simplification method \cite{garland1997surface}. For simplicity, the simplified 2-cycle is still denoted as $c$ and is treated as the \textit{initial control mesh} for Loop subdivision. The vertices within this control mesh are defined as \textit{control vertices}, still denoted by $V_c = \{v_1, \dots, v_n\}$.
After $k$ iterations of Loop subdivision, a refined mesh is generated, which exhibits improved quality and offers a more precise representation of the surface in the original point cloud. And we define the vertices in final refined mesh as \textit{mesh vertices}, denoted by $V_m^{(k)}=\{ v^{(k)}_1,\cdots, v^{(k)}_m\}$. It is worth noticing that each new mesh vertex in the refined mesh can be represented by a combination of control vertices according to the subdivision rules \cite{loop1987smooth}:
$$v^{(k)}_j=\sum_{i=1}^n a_i v_i,\: \text{with} \sum_{i=1}^n a_i=1,$$
where the coefficients $\{a_i\}$ can be derived from the subdivision rules.

To perform LSPIA, for each point $p_j\in P'$, we first compute its closest mesh vertex $v^{(k)} \in V_m^{(k)}$ and calculate the difference vector 
$$\delta_j = p_j-v^{(k)}.$$
Since $v^{(k)}$ can be represented by the combination of $V_{c}=\{ v_1,\cdots, v_n\}$, i.e. $$v^{(k)}=\sum_{i=1}^n a_{j,i} v_i.$$ 
The difference vector equals
$$\begin{aligned} 
	\delta_j =& p_j-(a_{j,1} v_1+\cdots + a_{j,n} v_n)\\
	=&a_{j,1} (p_j-v_1) +\cdots + a_{j,n} (p_j-v_n). 
\end{aligned}$$
Hence, denote $\delta_{j,i}:=(p_j-v_i)$, we have 
\begin{align}
	\delta_j=&a_{j,1} \delta_{j,1} +\cdots + a_{j,n} \delta_{j,n}.\label{eq:5}
\end{align}
Then the difference vector $\Delta_i$ distributed to the $i$-th control vertex $v_i$ is defined as
$$\Delta_i = \frac{\sum_j a_{j,i} \delta_{j,i}}{\sum_j a_{j,i}}.$$
Finally, adding this $\Delta_i$ to the control vertex $v_i$, we get the $i$-th new control vertex $v_i^{new}=v_i+\Delta_i$. By applying this to all control vertices, we obtained a new control mesh after one iteration.

The fitting steps described above are iteratively performed until the root mean square (RMS) fitting error at the 
$k$-th iteration, defined as 
$$e^{(k)}=\sqrt{\frac{\sum_{j=1}^{m_c} \|\delta_j^{(k)}\|^2}{{m_c}}},
$$
satisfies the stopping criterion 
$$\left | \frac{e^{(k+1)}}{e^{(k)}}-1\right |  <\varepsilon,$$
where $m_c$ is the number of points in $P'_c$, and $\varepsilon$ is a predefined threshold, typically set to $10^{-3}$. And the maximum number of iterations in our study is set to 100.

\begin{figure*}
	\centering
	\includegraphics[width=0.75\linewidth]{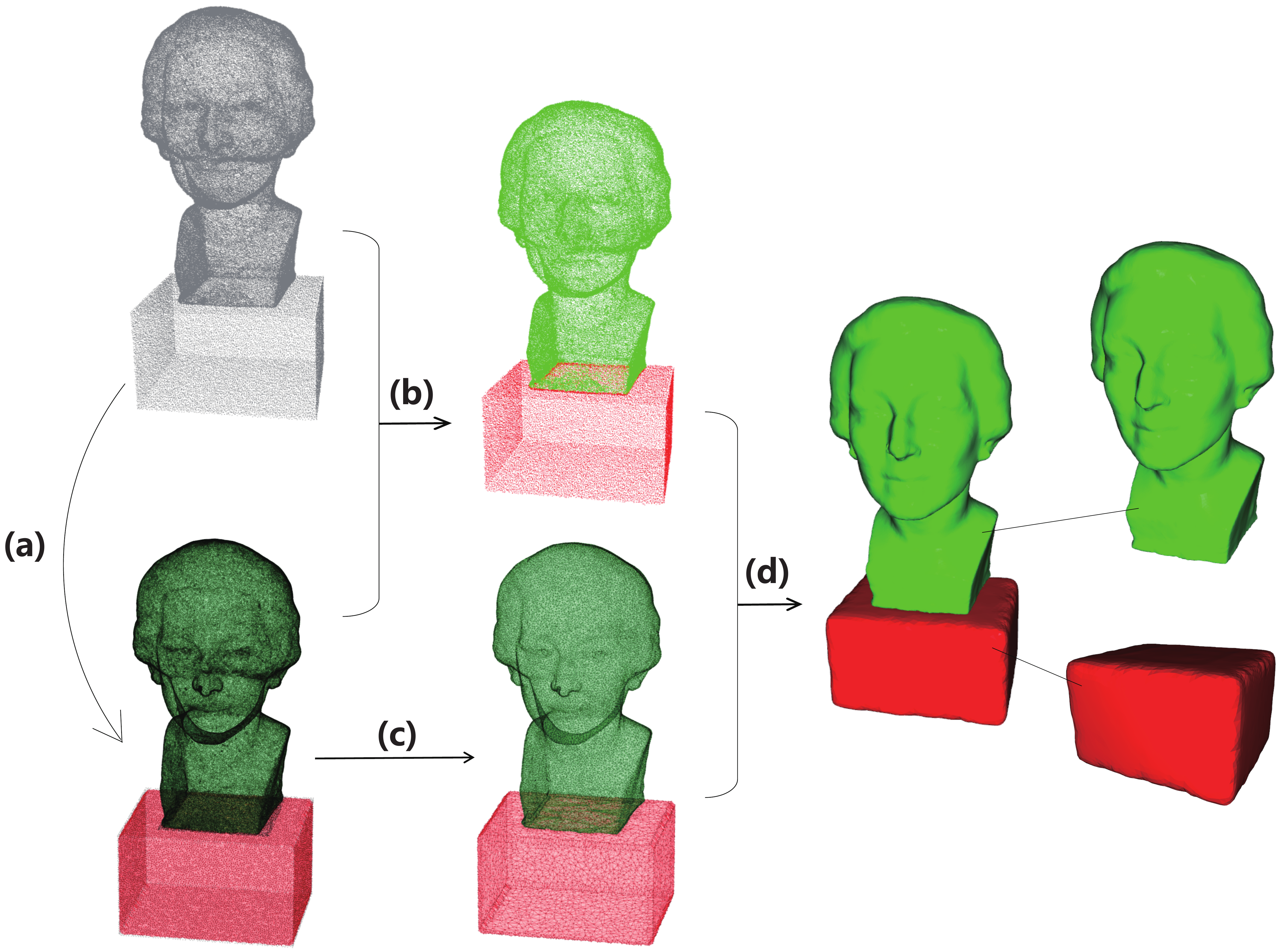}
	\caption{Illustration of the complete process of reconstructing a model from a noisy point cloud. (a) Compute representative 2-cycles of persistent homology from the point cloud. (b) Determine the neighboring point cloud from the original point cloud for each 2-cycle by $d_{avg}$, displayed in different colors. (c) Reduce the number of meshes in 2-cycles to get initial control meshes. (d) The reconstructed model, whose components are also depicted.}
	\label{fig:illustration}
\end{figure*}

Therefore, by utilizing the above operations for each significant 2-cycle, we can finally get the reconstruction results.
In conclusion, the outline of the proposed algorithm for model reconstruction is listed as follows:
\begin{algorithm}[H]  
	\renewcommand{\algorithmicrequire}{\textbf{Input:}}    
	\renewcommand{\algorithmicensure}{\textbf{Return:}}    
	\caption{Model Reconstruction Algorithm}  
	\label{alg:sr}  
	\begin{algorithmic}[1]  
		\REQUIRE A point cloud $P$ representing a model.  
		\STATE Construct the alpha filtration from $P$, then calculate persistent homology and the 2-PD. 
		\STATE Cluster the points in 2-PD and derive the significant points.
		\STATE Compute persistent volume and volume-optimal cycle for each significant point in 2-PD.
		\STATE Identify all non-manifold vertices and edges for each volume-optimal cycle, then remove 3-simplices in the corresponding persistent volume that connect with non-manifold vertices or edges. Compute the boundary of the final persistent volume as the new representative 2-cycle.
		\STATE Compute neighboring points (Eq \ref{eq:4}) for each 2-cycle.
		\STATE Reduce each 2-cycle to a simpler mesh using the QEM method, then use the reduced 2-cycle as the initial control mesh and apply Loop subdivision to generate refined meshes.
		\STATE Apply LSPIA to optimize each obtained mesh surface with its neighboring points until the stopping criterion is reached.
		\ENSURE The reconstructed surfaces of the components of the model. 
	\end{algorithmic}  
\end{algorithm} 

Now, we analyze the computational complexity of Algorithm \ref{alg:sr}. Let $n$ denote the number of points in the point cloud, $m$ the total number of points in the derived 2-PD, and $k$ the number of significant points in the 2-PD (hence $k < m$). The primary computational cost in Step 1 arises from constructing the alpha filtration, which has a complexity of $O(n \log n+K)$ since the alpha complex is derived from the Delaunay triangulation, and here $K$ is the size of the Delaunay triangulation. Step 2, the 2-means clustering, has a computational complexity of $O(m)$. Steps 3 to 7 are repeated for each of the $k$ significant points. For the $i$-th iteration ($1 \leq i \leq k$), let $N_i$ denote the number of 3-simplices in the corresponding persistent volume, and $V_i$ the number of vertices in the representative cycle. Specifically, Step 3 has a complexity of $O(N_i)$ \cite{obayashi2018volume}; Step 4 has a complexity of $O(N_i)$ for sufficient sampling in practice; Step 5 achieves $O(n \log n)$ complexity using a KD-tree; the QEM mesh simplification in Step 6 incurs $O(V_i \log V_i)$ complexity; and according to Subsection \ref{LSPIA}, the LSPIA fitting in Step 7 has complexity $O(n V_i)$. Given that $k$ can be regarded as a constant, and each $V_i$ satisfies $V_i \leq n$, $N_i=O(V_i^2)$, the overall computational complexity of Algorithm \ref{alg:sr} is approximately $O(n^2)$.

\begin{figure*}
	\centering
	\includegraphics[width=0.8\linewidth]{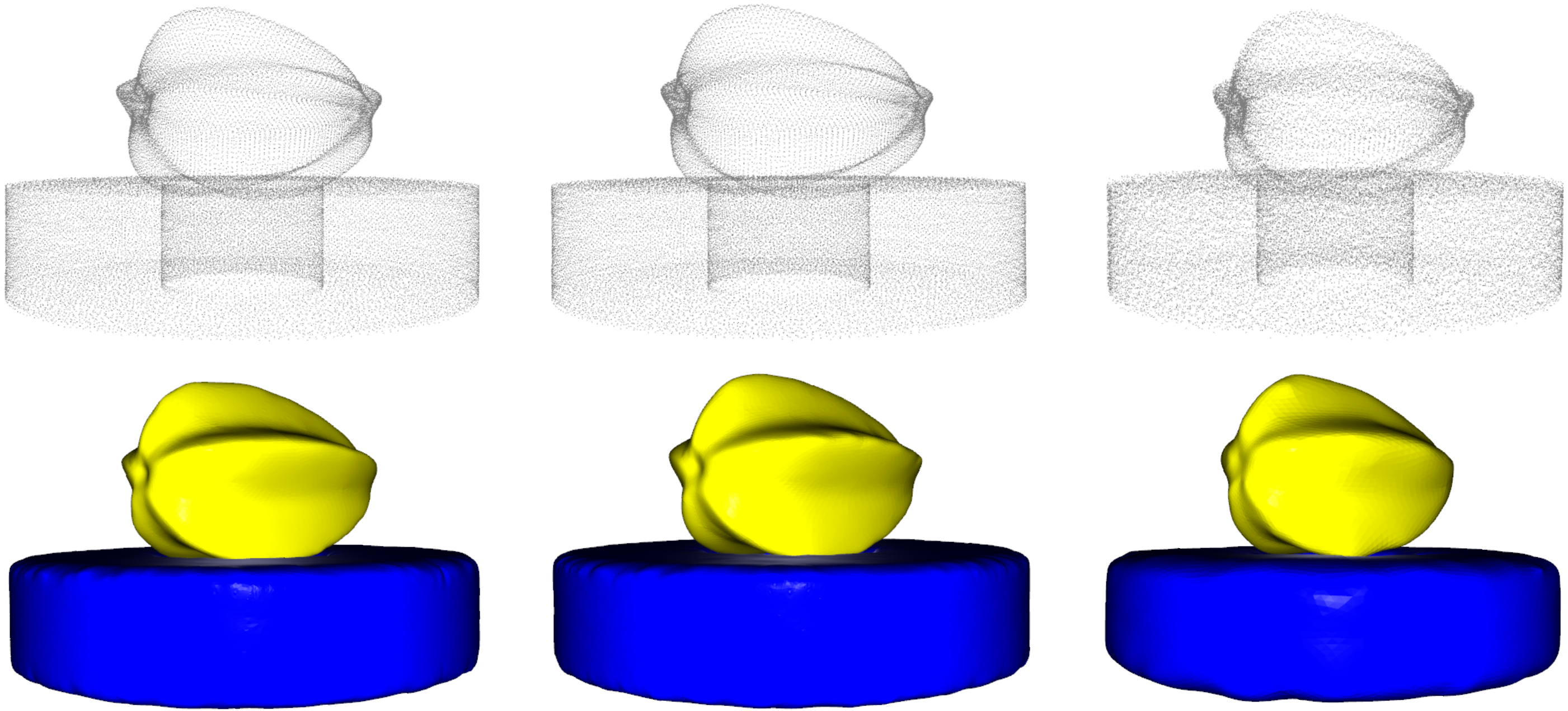}
	\caption{Illustration of the robustness of our method to noise. Upper: from left to right, the noise of the point cloud increases. Bottom: the corresponding reconstructed models using our method.}
	\label{fig:noise}
\end{figure*}

\begin{figure*}
	\centering
	\includegraphics[width=1.0\linewidth]{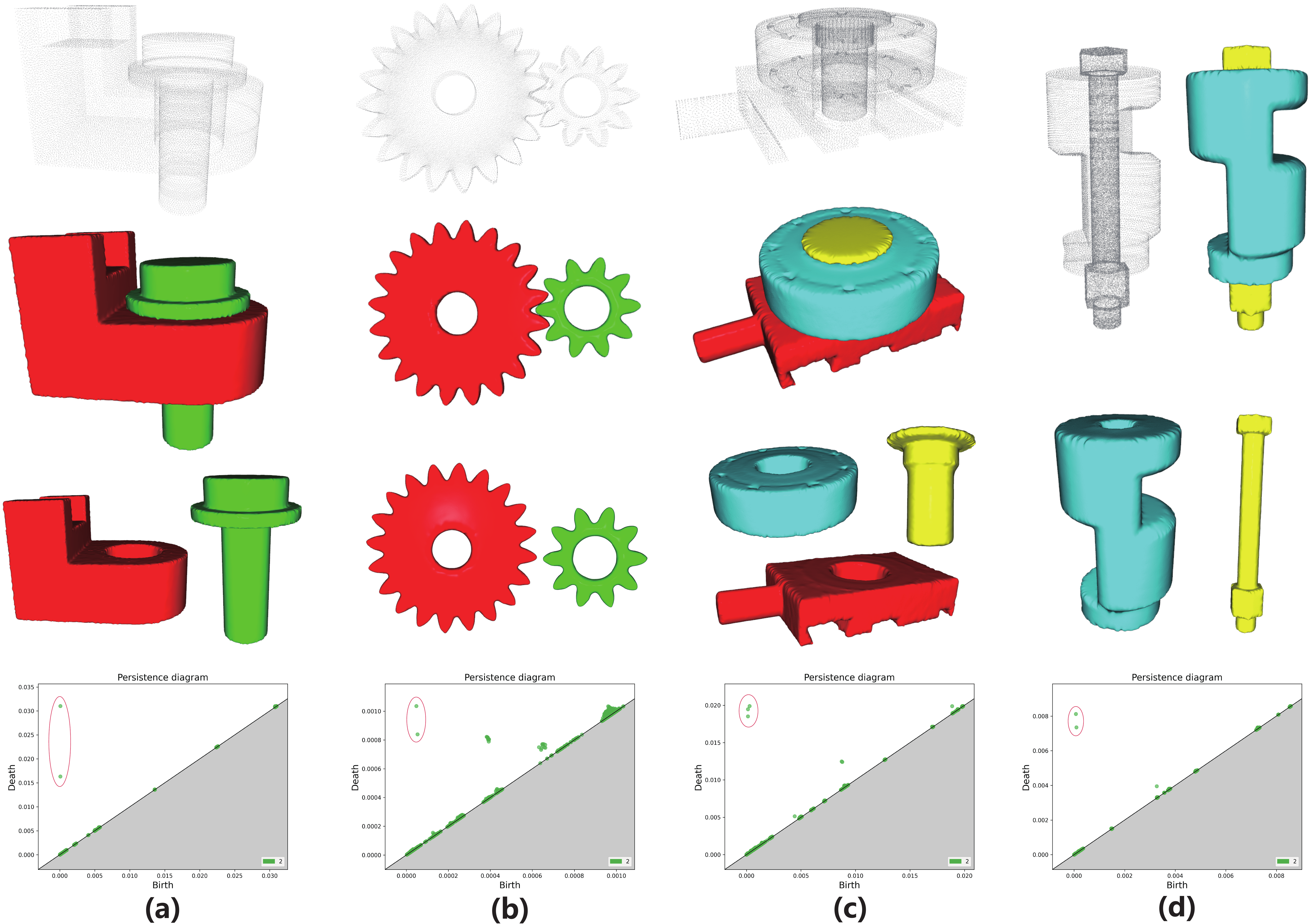}
	\caption{Experimental results of reconstructed mechanical models composed of multiple mechanical components commonly found in CAD.}
	\label{fig:experiments1}
\end{figure*}

\begin{table*}[h!t]
	\centering  
	\caption{Statistics on point clouds and reconstructed results.}
	\label{tab:RMS}  
	\resizebox{0.9\textwidth}{!}{%
		\begin{tabular}{|c|c|c|c|c|c|c|}  
			\hline  
			Point Cloud
			& $\begin{array}{c}
				\text { Number of } \\
				\text { Points }
			\end{array}$ 
			& $\begin{array}{c}
				\text { Number of } \\
				\text { Control Vertices }
			\end{array}$
			& $\begin{array}{c}
				\text { Number of } \\
				\text { Surfaces }
			\end{array}$
			& RMS 
			& $\begin{array}{c}
				\text { Time of } \\
				\text { Computing Topology }
			\end{array}$
			& $\begin{array}{c}
				\text { Time of } \\
				\text { Subdivision Fitting }
			\end{array}$\\
			\hline
			Fig. \ref{fig:robothand} & 84157 & 31314 & 5 & $ 3.277\times 10^{-4}$ &  1506.5s & 463.7s \\ 
			\hline
			Fig. \ref{fig:experiments1} (a) & 34173 & 8332 &2 & $1.184\times 10^{-4}$ & 239.5s & 520.0s \\ 
			\hline  
			Fig. \ref{fig:experiments1} (b) & 38644 & 3452 &2 & $1.476\times 10^{-5}$ & 195.8s & 466.0s\\  
			\hline 
			Fig. \ref{fig:experiments1} (c) & 45618 & 14870 &3 & $ 3.339\times 10^{-4}$ & 404.0s & 684.2s\\  
			\hline 
			Fig. \ref{fig:experiments1} (d) & 50146 & 12163 &2 & $1.649\times 10^{-4}$ & 351.6s & 925.8s\\ 
			\hline
			Fig. \ref{fig:experiments2} (a) & 300869 & 51692 &2 & $2.246\times 10^{-4}$ & 22122.0s & 1384.9s\\ 
			\hline  
			Fig. \ref{fig:experiments2} (b) & 598144 & 67050 &2 & $ 7.178\times 10^{-4}$ & 16987.1s & 2822.7s\\  	
			\hline
			Fig. \ref{fig:experiments2} (c) & 87656 & 9986 &2 & $1.828\times 10^{-4}$ & 1454.9s & 523.1s\\ 
			\hline  
			Fig. \ref{fig:experiments2} (d) & 73312 & 6943 &4 & $ 1.178\times 10^{-4}$ & 1591.8s & 421.2s\\  	
			\hline
		\end{tabular}%
	}  
\end{table*} 
\section{Experiments and Discussions}\label{sec:exp}
In this section, the experimental results of the proposed
method are presented. All the experiments were conducted on a server equipped with dual Intel(R) Xeon(R) CPU E5-2678 v3 processors (2.50GHz, 24 cores, 48 threads) and 62 GB of RAM. In addition, comparisons with previous methods are provided. Finally, we will give some discussions.

\begin{figure*}
	\centering
	\includegraphics[width=1.0\linewidth]{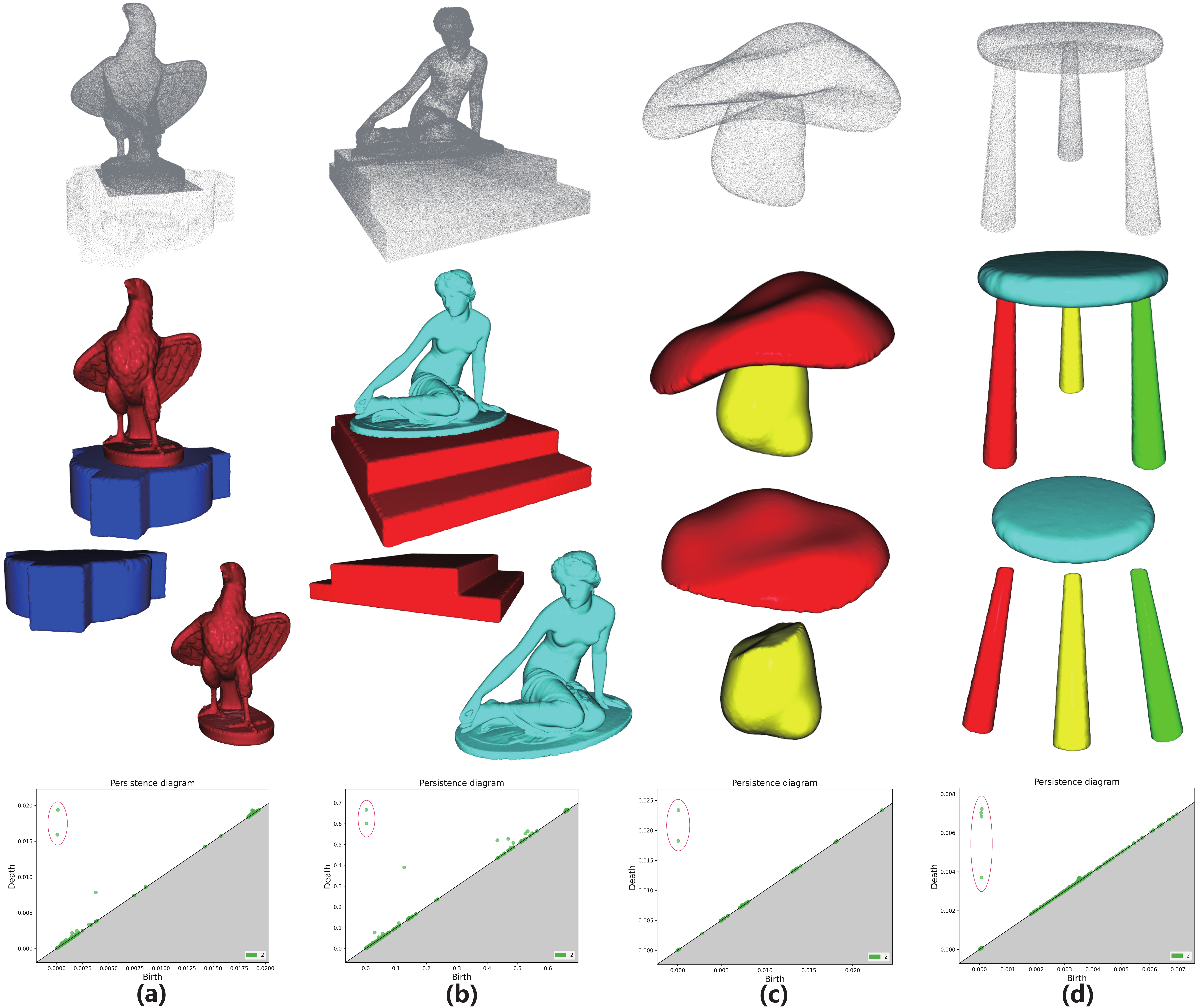}
	\caption{Experimental results of reconstructed models composed of renowned sculptures (a,b) or real-world objects (c,d).}
	\label{fig:experiments2}
\end{figure*}

\begin{figure*}
	\centering
	\includegraphics[width=0.95\linewidth]{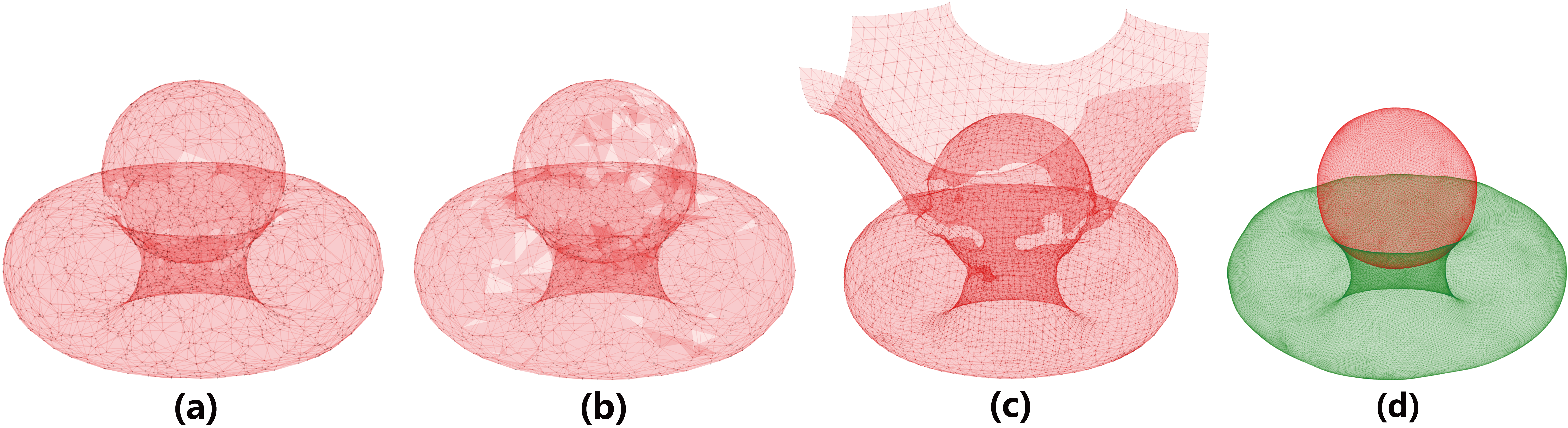}
	\caption{The comparison results. (a) Result of the greedy Delaunay method. (b) Result of the ball-pivoting method. (c) Result of the screened Poisson surface reconstruction method. (d) Result of our method.}
	\label{fig:compare}
\end{figure*}

\subsection{Experimental results}
We begin by demonstrating the whole process of reconstructing models using our approach. In Fig. \ref{fig:illustration}, the input point cloud represents a sculpture and its base. We compute the topological representative 2-cycles (Fig. \ref{fig:illustration} a) from the point cloud and remove non-manifold vertices and edges, then determine the neighboring point cloud (Eq \ref{eq:5}) of each representative cycle using $d_{avg}$ (Fig. \ref{fig:illustration} b). Subsequently, the initial control meshes are determined by calculating the reduced 2-cycles (Fig. \ref{fig:illustration} c). Finally, the Loop subdivision fitting is utilized to obtain the final reconstructed models (Fig. \ref{fig:illustration} d).

Additionally, our method is robust to noise in point clouds. In Fig. \ref{fig:noise}, the upper figures illustrate the progressive addition of noise to the original point cloud, while the bottom figures depict the corresponding reconstruction results of the proposed method. It is clear that our approach successfully identifies and reconstructs all closed surfaces for each given model, which demonstrates that our method is robust to noise.

Moreover, we show that the proposed approach effectively addresses key challenges in reconstructing models with multiple surface components. Specifically, we showcase the reconstruction of closed surfaces with shared regions, emphasizing the unique advantages and efficacy of our method. 
For example, for the model illustrated in Fig. \ref{fig:robothand}, our method successfully decomposes and reconstructs the surface of each component with an accurate shape. This demonstrates that our method has strong potential for applications in reverse engineering. 
Fig. \ref{fig:experiments1} and \ref{fig:experiments2} present more experimental results of reconstructed models using our method. Here, our examples comprise point clouds of CAD model surfaces from ABC dataset \footnote{\url{https://deep-geometry.github.io/abc-dataset}} \cite{koch2019abc}, real-world object models from 3DNet dataset \footnote{\url{https://strands.readthedocs.io/en/latest/datasets/three_d_net.html}} \cite{wohlkinger20123dnet}, and real-world renowned sculptures \footnote{\url{https://threedscans.com}}. 
Fig. \ref{fig:experiments1} shows several models made up of multiple mechanical components commonly seen in CAD. 
Fig. \ref{fig:experiments2} (a--b) shows some sculptures and their bases, which are typical in computer graphics. 
Fig. \ref{fig:experiments2} (c--d) shows real-world object models (mushroom and stool).
Each input point cloud represents a model containing multiple closed surfaces with shared regions and includes the presence of noise, and each reconstructed closed surface is drawn with a different color.
Additionally, the 2-PDs derived from each example are presented beneath these figures, with the significant points identified through clustering clearly highlighted. These significant points exhibit substantially greater persistence compared to noise points, as evidenced by their larger distances to the diagonal. In our examples, each distinct component corresponds to precisely one significant point in the 2-PD.

Table \ref{tab:RMS} summarizes the experimental results, including the number of points in each point cloud, the total number of different control vertices of all control meshes, the number of obtained closed surfaces, the final root mean square (RMS) error for all reconstructed surfaces, the running time of computing persistent homology and representative cycles, and the time of performing subdivision fitting with LSPIA. The final RMS is calculated using the formula
\begin{align}
	RMS = \sqrt{\frac{\sum_{j=1}^m \|\delta_j\|^2}{m}},
\end{align}
with $m$ the number of points in the neighboring point cloud (Eq \ref{eq:4}) and $\delta_j$ defined as (Eq \ref{eq:5}). We can see that the proposed method has demonstrated relatively small RMS on the test point clouds, showcasing its effectiveness in model reconstruction.

\subsection{Comparison with other methods}
We compared our method with classical surface reconstruction techniques, including the greedy Delaunay method \cite{cohen2004greedy} (results are computed by CGAL \cite{fabri2009cgal}), the ball-pivoting method \cite{bernardini1999ball} (results are computed by MeshLab \cite{cignoni2008meshlab}), and the screened Poisson surface reconstruction method \cite{kazhdan2013screened} (results are computed by MeshLab).
Fig. \ref{fig:compare} presents the comparison results of the reconstruction results of a point cloud representing a tangent sphere and torus. As shown, both the greedy Delaunay method and the ball-pivoting method (Fig. \ref{fig:compare} (a) and (b)) fail to completely reconstruct the shared regions between the sphere and the torus, resulting in incomplete meshes. The screened Poisson surface reconstruction method (Fig. \ref{fig:compare} (c)) generates excessive redundant meshes due to inaccuracies in the normal computations at the shared regions.
More importantly, none of these methods successfully separate the sphere and the torus from the point cloud, while our method (Fig. \ref{fig:compare} (d)) effectively identifies and reconstructs them. These comparisons demonstrate that our method outperforms the others in handling models with multiple components and preserving topological features.

\subsection{Discussions}
\begin{figure*}
	\centering
	\includegraphics[width=0.75\linewidth]{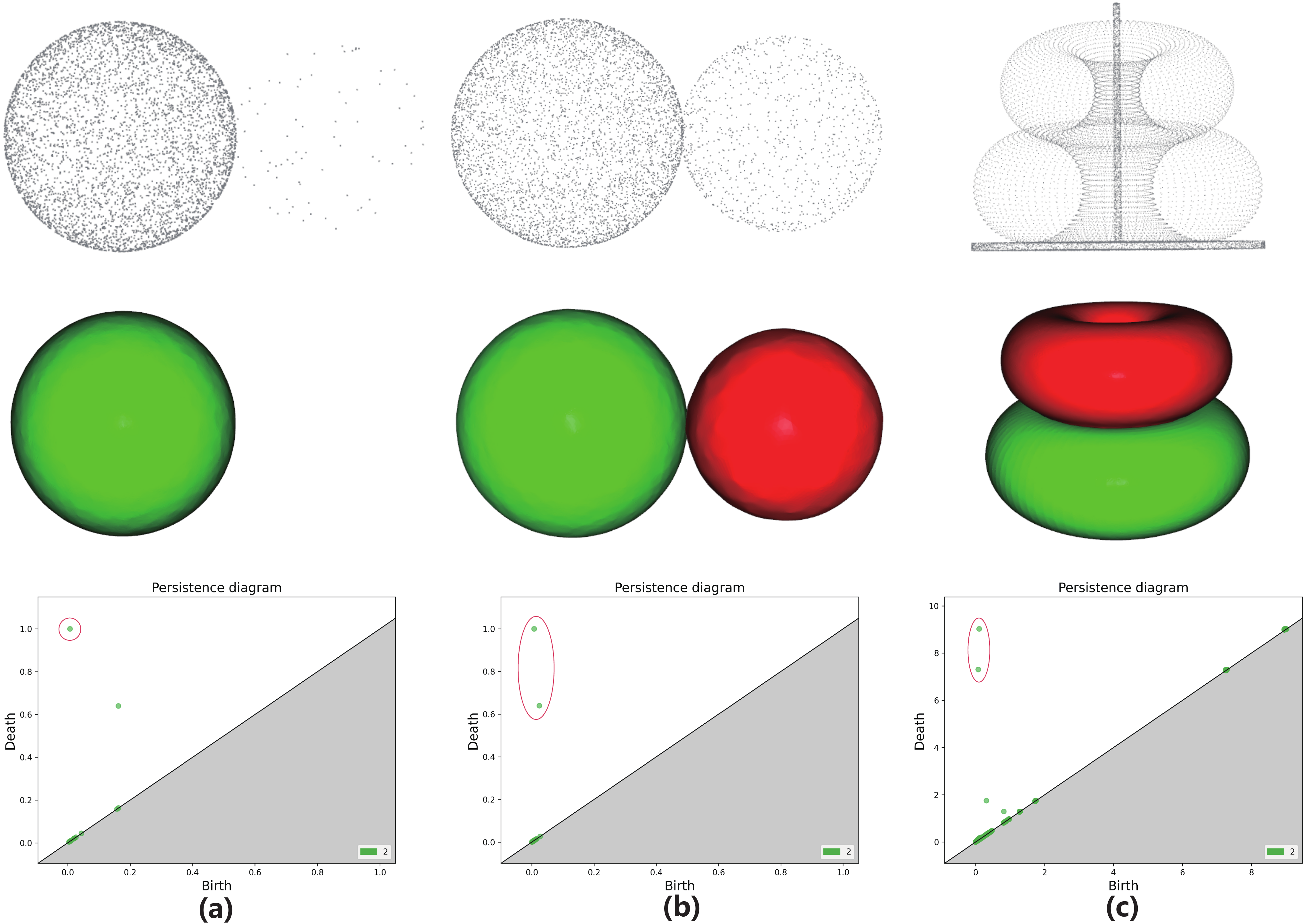}
	\caption{Illustration of some limitations. (a) Successfully reconstruct the left ball but fail to reconstruct the right ball because under-sampling makes the corresponding feature become insignificant. (b) After obtaining sufficient sampling points, both of the two balls in (a) can be successfully reconstructed. (c) Fail to reconstruct thin or narrow components due to missing features that should be identified as significant.}
	\label{fig:limitations}
\end{figure*}
\begin{figure}
	\centering
	\includegraphics[width=0.8\linewidth]{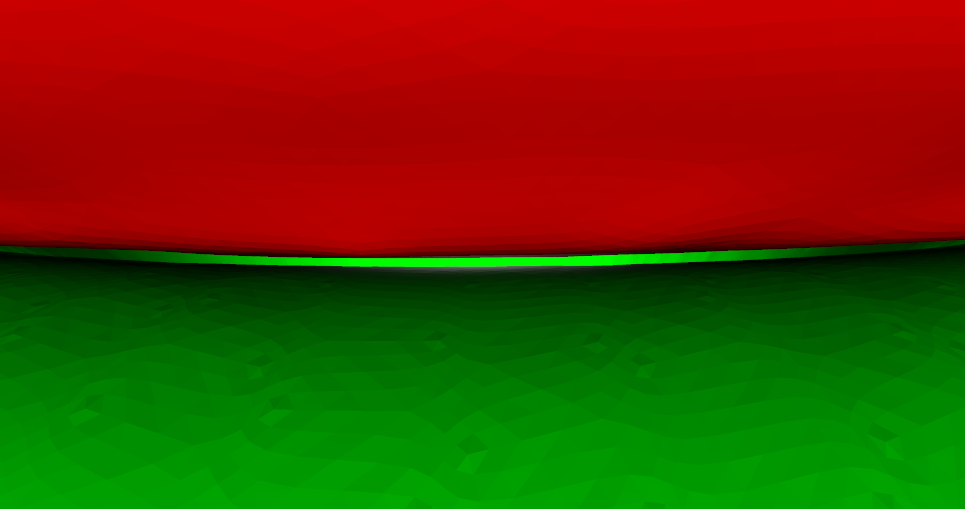}
	\caption{Illustration of gap between components in Fig. \ref{fig:limitations} (c).}
	\label{fig:incompatible}
\end{figure}

In our method, the persistence diagram relies on a sufficiently dense sampling point cloud to accurately capture topological information. When sampling density is insufficient, the PD may contain fewer significant points than the number of closed surfaces we aim to reconstruct, since an insufficient sample will reduce the persistence of corresponding 2-cycles by delaying their birth time. Fig. \ref{fig:limitations} demonstrates an example, (a) shows failure to reconstruct an undersampled sphere, while (b) demonstrates successful reconstruction after sufficient sampling. Additionally, topological features corresponding to thin, narrow, or small components often exhibit low persistence values in the PD, making them difficult to distinguish when larger components are present. An example is illustrated in Fig. \ref{fig:limitations} (c), where two thin or narrow components fail to be reconstructed. The reconstruction process also faces challenges with adjacent components: the Loop subdivision fitting may lead to gaps between components in the reconstructed models (see Fig. \ref{fig:incompatible} for an instance, which illustrates the gap between two surfaces in Fig. \ref{fig:limitations} (c)), especially under high noise conditions. This limitation may affect downstream applications requiring precise interface identification between components. These constraints constitute important considerations for our future research.

In addition, our method specializes in reconstructing closed surfaces with an emphasis on topology and global shape preservation. While it effectively handles smooth surfaces with shared regions (e.g., industrial mechanisms and sculptures), it does not guarantee the preservation of sharp geometric features such as sharp corners. This is because the Loop subdivision and LSPIA optimization steps inherently smooth the surface to achieve high-quality meshes, which may inadvertently blur such geometric details. Preserving more geometric details is also a valuable future work.

\section{Conclusions and Future Work}\label{sec:conclusion}
In this paper, we develop a novel method for reconstructing models with multiple components, enabling the reconstruction of complete models composed of multiple components represented by point clouds. By integrating persistent homology with Loop subdivision fitting and LSPIA, our method effectively identifies and separates multiple closed surfaces in a model, including those with shared regions, and generates high-quality reconstructions that are robust to noise. Experimental results demonstrate the efficacy of our approach and underscore its potential for practical applications. 

In future work, we aim to further explore methods of alignment between common surfaces of components in reconstruction models (e.g., using the idea of material interfaces \cite{du2022robust}) for downstream applications. Additionally, we will further explore the topological understanding provided by persistent homology to obtain more accurate and comprehensive topological information from the given point clouds.

\section*{Acknowledgment}

\bibliographystyle{elsarticle-num}
\bibliography{BIB.bib}

\end{document}